\documentclass[]{aa} 
\usepackage{amsmath}
\usepackage{graphicx}
\usepackage{txfonts}
\usepackage{url}
\usepackage{bm}

\date{\today}

\title{Interpretation of increased energetic particle flux measurements by SEPT aboard the STEREO spacecraft and contamination}

\titlerunning{SEPT particle observations during the CIR event on 9 August 2011}
\authorrunning{Wraase et al.}

\author{Wraase, S.\inst{1}
\and Heber, B.\inst{1}
\and B\"ottcher, S. \inst{1}
\and Bucik, R. \inst{3}
\and Dresing, N. \inst{1}
\and G\'omez-Herrero\inst{2}
\and Klassen, A. \inst{1}
\and M\"uller-Mellin, R.\inst{1}
}

\institute{Institut f\"ur Experimentelle und Angewandte Physik, Christian-Albrechts-Universit\"at zu Kiel, Leibnizstraße 11, D-24118 Kiel, Germany, \email{heber@physik.uni-kiel.de}
\and Dpto. de F\'isica y Matem\'aticas, Universidad de Alcal\'a, E-28871 Alcal\'a de Henares, Madrid, Spain
\and Georg-August-Universit\"at G\"ottingen, Institut f\"ur Astrophysik,Friedrich-Hund-Platz 1,D-37077 G\"ottingen. Germany
}

\abstract{
Interplanetary (IP) shocks are known to be accelerators of energetic charged particles observed in-situ in the heliosphere. However, the acceleration of near-relativistic electrons by shocks in the interplanetary medium is often questioned. On 9 August 2011 a Corotating Interaction Region (CIR) passed STEREO B (STB) that resulted in a flux increase in the electron and ion channels of the Solar Electron and Proton Telescope (SEPT). Because electron measurements in the few~keV to several 100~keV range rely on the so-called magnet foil technique, which is utilized by SEPT, ions can contribute to the electron channels.}
{We aim to investigate whether the flux increase in the electron channels of SEPT during the CIR event on 9 August 2011 is caused by ion contamination only.}
{We compute the SEPT response functions for protons and helium utilizing an updated GEANT4 model of SEPT. The CIR energetic particle ion spectra for protons and helium are assumed to follow a Band function in energy per nucleon with a constant helium to proton ratio.}
{Our analysis leads to a helium to proton ratio of 16.9~\% and a proton flux following a Band function with the parameters $I_0=1.24 \cdot 10^4$~/~(cm$^2$~s~sr~MeV/nuc.), $E_c=79$~keV/nuc. and spectral indices of $\gamma_1=-0.94$ and $\gamma_2=-3.80$ which are in good agreement with measurements by the Suprathermal Ion Telescope (SIT) aboard STB.}
{Since our results explain the SEPT measurements, we conclude that no significant amount of electrons were accelerated between 55~keV and 425~keV by the CIR.}

\keywords{Energetic Particle Measurements - Magnet Foil Technique - Ion Contamination - Corotating Interaction Region}

\begin{document}
\maketitle

\section{Introduction}
The two Solar TErrestrial RElations Observatory \citep[STEREO,][]{Kaiser-etal-2008} spacecraft were launched in October 2006 in the declining phase of solar cycle 23. During the first years of the mission the observed particle enhancements were mainly dominated by recurrent energetic particle events associated with Corotating Interaction Regions \citep[CIRs, ][and references therein]{Bucik-etal-2012,Dresing-etal-2009,Leske-etal-2008}. CIRs are large-scale plasma structures generated in low and middle latitude regions of the heliosphere by the interaction between a fast solar wind stream, with typical speeds of about 800~km/s, and the surrounding slow solar wind, with speeds of about 400~km/s. 
If the source regions of the solar wind streams at the Sun are stable, a stationary observer at e.g. 1~au will note recurrent fast and slow solar wind streams with a period of one solar rotation. If (1) the pressure gradient becomes sufficiently strong and (2) the speed difference of the streams exceeds the local magnetosonic speed, shocks can be formed at the edges of the compression region of the structure. Because these shocks are moving quasi-radially outward and the magnetic field follows the Archimedian spiral they are quasi-perpendicular.
An observer close to the ecliptic will measure at these distances first a Forward Shock (FS), when the speed and the magnetic field strength increase, second the Stream Interface (SI), which is the surface along the two solar wind streams interacting with each other, and a Reverse Shock (RS) with typical signatures of an increase of the solar wind speed and a decrease of magnetic field strength. 
These shocks are known to accelerate ions up to several MeV per nucleon. In these cases the energy per nucleon spectrum is found to be nearly the same for most ions like protons and helium with an averaged helium to proton ratio of more than 10~\%. While ion increases have been frequently observed in association with CIR shocks \citep{Barnes-Simpson-1976,McDonald-etal-1976,Heber-etal-1999}, electrons often show different time profiles \citep{Kunow-etal-1999}. For a review of the effects of CIRs on energetic particles, see \citet{Richardson-2004} and references therein. 

\section{Observations}
\begin{figure}
\centering
\includegraphics[width=\columnwidth]{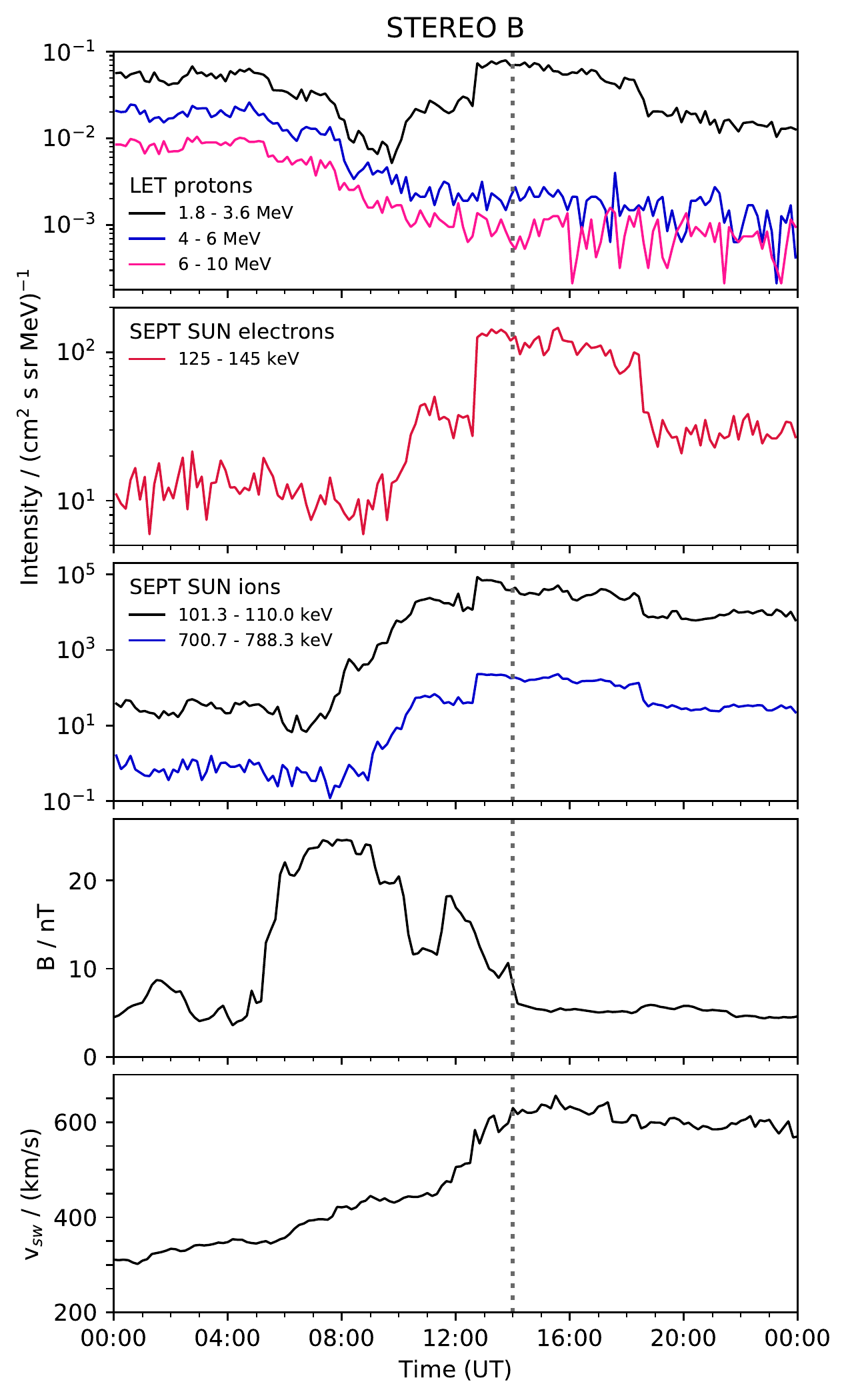}
\caption{From top to bottom: Ten minutes averaged fluxes of 1.8--3.6~MeV (black curve), 4--6~MeV (blue curve) and 6--10~MeV protons (pink curve), 125--145~keV electrons (red curve), 100--110~keV (black curve) and 700--780~keV (blue curve) protons, magnetic field strength and solar wind speed on 9~August 2011 observed by STB. The dashed line marks the time of the RS (for details see text).}
\label{fig:1}
\end{figure}

\begin{figure}
\includegraphics[width=\columnwidth]{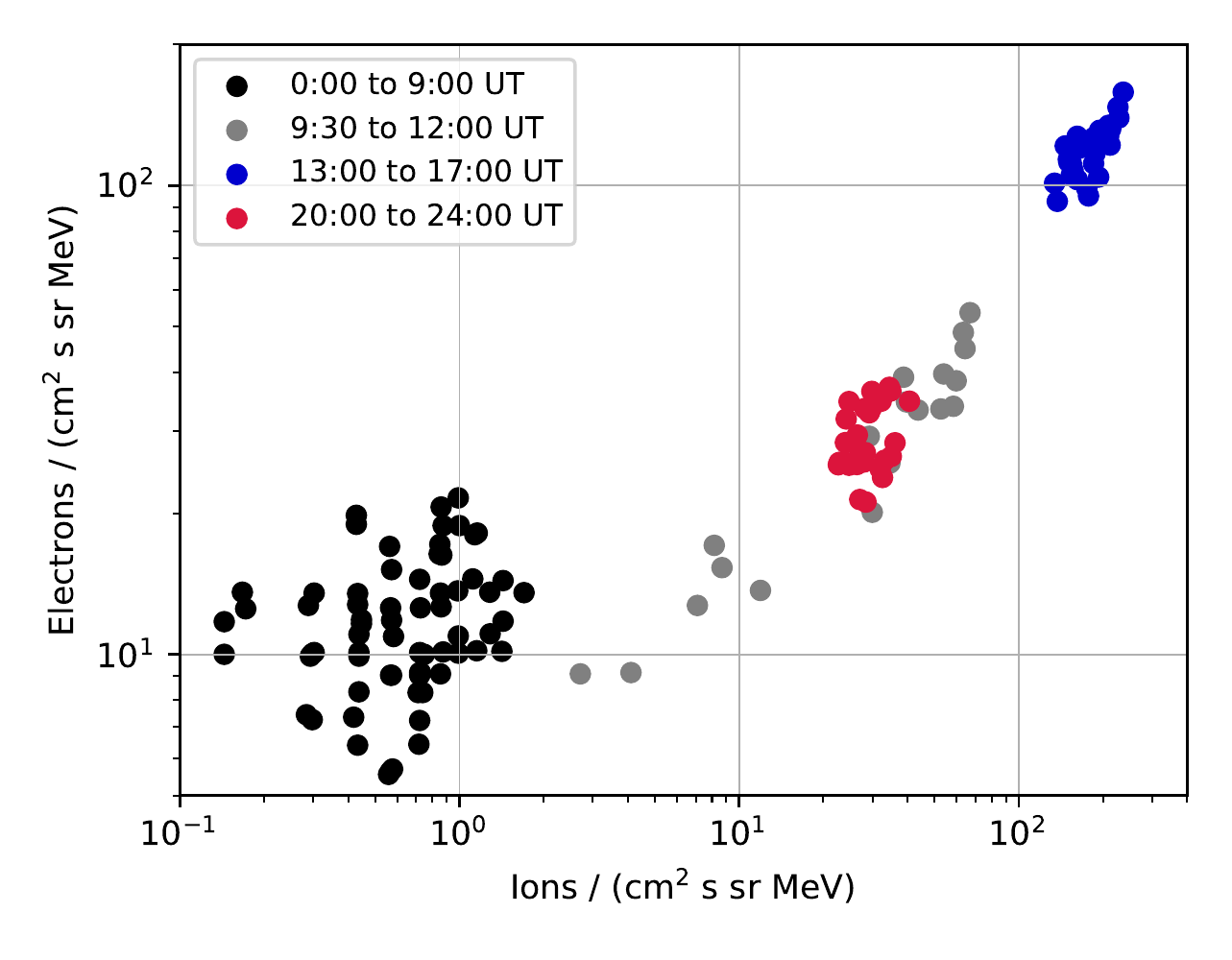}
\caption{SEPT 125--145~keV electron flux (second panel in Fig.~\ref{fig:1}) as function of the 700--780~keV ion flux (blue curve in the third panel of Fig.~\ref{fig:1}). Marked by different colors are the pre-event and different phases of the recurrent ion event. While no correlation is seen for the SEPT measurements when the ion flux is smaller than $5$~ions/(cm$^2$~s~sr~MeV) a clear dependence is shown for the ions above that threshold.}
\label{fig:2}
\end{figure}

During 9~August 2011 a RS with a magnetosonic Mach number of one  was measured by STEREO B (STB) at 13:58 UT (http://www.ipshocks.fi) as indicated by the dashed line in Fig.~\ref{fig:1}. This figure displays from top to bottom the omni-directional time history of 1.8--3.6~MeV (black curve), 4--6~MeV (blue curve) and 6--10~MeV protons \citep[measured by the Low Energy Telescope,][]{Mewaldt-etal-2008}, 125--145~keV electrons (red line), 100--110~keV (black line) and 700--780~keV (blue line) ions measured by the SUN-telescope of the Solar Electron and Proton Telescope \citep[SEPT,][]{Mueller-Mellin-etal-2008}, the magnetic field strength, and the solar wind speed measured by the STEREO Magnetic Field Experiment \citep[MAG,][]{Acuna-etal-2008} and STEREO Plasma and Suprathermal Ion Composition \citep[PLASTIC,][]{Galvin-etal-2008}, respectively. Since no forward shock was reported the CIR structure was not fully developed when passing the spacecraft from about 5 to 19~UT. As can be seen in Fig.~\ref{fig:1} the flux of 100--110~keV ions shows a maximum intensity during the passage of the RS, indicating that these low energy ions are accelerated by the shock. Similar time profiles are measured by SEPT at higher energies, too (blue curve in middle panel of Fig.~\ref{fig:1}). Proton measurements above 4~MeV by the Low Energy Telescope (LET) show no intensity increase during the CIR passage. The 125--145~keV electron intensity time profile, shown in the second panel of Fig.~\ref{fig:1}, follows the 700~keV proton intensity time profiles from 10 UT onwards. We would like to point out that \citet{Xu-etala-2017} interpreted this electron increase measured by SEPT as a signature of the solar energetic particle event caused by an X6.9 solar flare accompanied by a halo coronal mass ejection. However, Fig.~\ref{fig:2} indicates a close correlation of SEPT ions and electrons  because the 125--145~keV electron flux depends on the 700--780~keV ion flux for periods from 10 to 24 UT on 9 August 2011.

Fig.~\ref{fig:2} indicates two energetic particle populations: Values in the lower left corner correspond to  measurements in time until 9~UT (see Fig.~\ref{fig:1}). Values above $5$~ions/(cm$^2$~s~sr~MeV) are obtained during different phases of the recurrent ion event. The maximum values are found from about 13 to 17~UT, a period that includes the spacecraft crossing of the reverse shock. While there is no obvious correlation for fluxes below 5~ions/(cm$^2$~s~sr~MeV) a clear correlation is found for all values above 10~ions/(cm$^2$~s~sr~MeV). However, a correlation does not exclude the possibility of a small contribution of electrons accelerated by the RS shock or  from electrons coming from the flare mentioned above. In order to examine the problem in detail we applied a GEANT4 simulation of SEPT to investigate the origin of the electron flux time profiles.

\section{Instrument Description and Analysis}
Measurements of electrons in the energy range of a few tens up to several hundreds of keV are based on the magnet foil technique. This technique is used by the Solar Electron and Proton Telescope \citep[SEPT,][]{Mueller-Mellin-etal-2008} aboard STEREO \citep{Kaiser-etal-2008}. In order to separate electrons from ions each sensor consists of a double-ended telescope with one end measuring primarily electrons and the other one ions.

\begin{figure}
\centering\includegraphics[width=0.92\columnwidth, trim=0cm 4cm 0cm 4cm, clip]{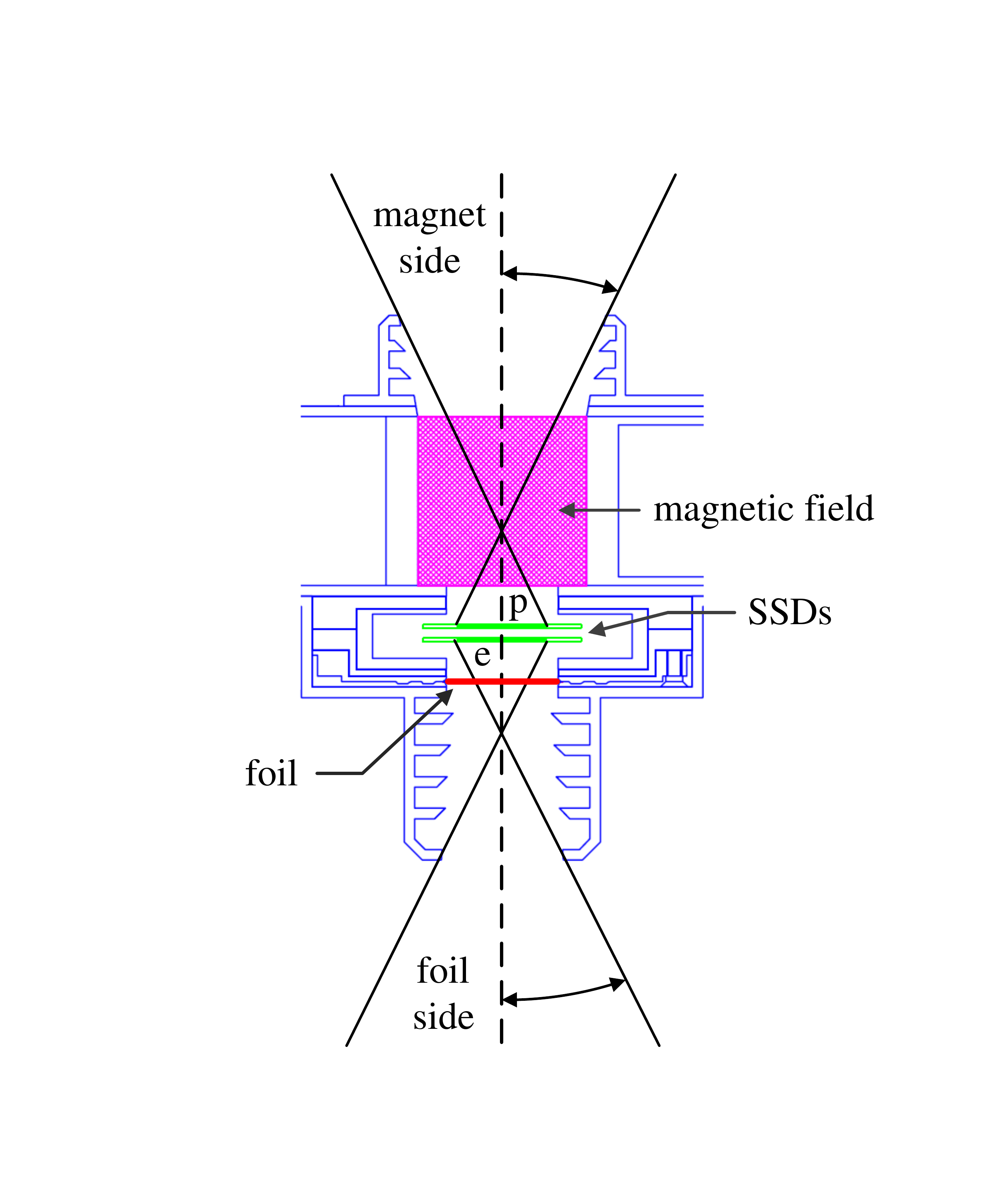}
\caption{Sketch of one of the SEPT double-ended telescopes consisting of two SSDs (green lines) with one looking through a foil (red line) and the other one through a magnetic field (magenta area).} 
\label{fig:sketch}
\end{figure}

As sketched in Fig.~\ref{fig:sketch} each SEPT telescope consists of two solid-state detectors (SSDs) that are operated in anti-coincidence. One SSD looks through an absorption foil and the other one through the gap of a magnet, hence through a magnetic field. The foil leaves the electron spectrum essentially unchanged but stops protons with energies up to the energy of electrons penetrating the SSD, $E_{max,e}$. The magnet is designed to sweep away electrons with $E<E_{max,e}$, but leaves ions unaffected. For each SSD particles are registered by their deposited energy in 32 bins ranging from 35~keV to 2.2~MeV for the magnet (ions) and foil side (electrons). Only in the absence of ions with $E>E_{max,e}$, the foil SSD detects only electrons, and in the absence of electrons with $E>E_{max,e}$, the magnet SSD only detects ions. 
The energy of ions needed to penetrate the SSD, $E_{max,p}$, determines the upper energy limit of ions stopping in the magnet SSD for which the energy spectrum can be measured quite cleanly. The contribution of $E>E_{max,e}$ ions to the foil SSD energy spectrum can then be computed and subtracted from the observed spectrum to obtain the pure electron spectrum. 

\begin{figure}
\includegraphics[width=\columnwidth, trim=0.5cm 1.2cm 0.cm 1cm, clip]{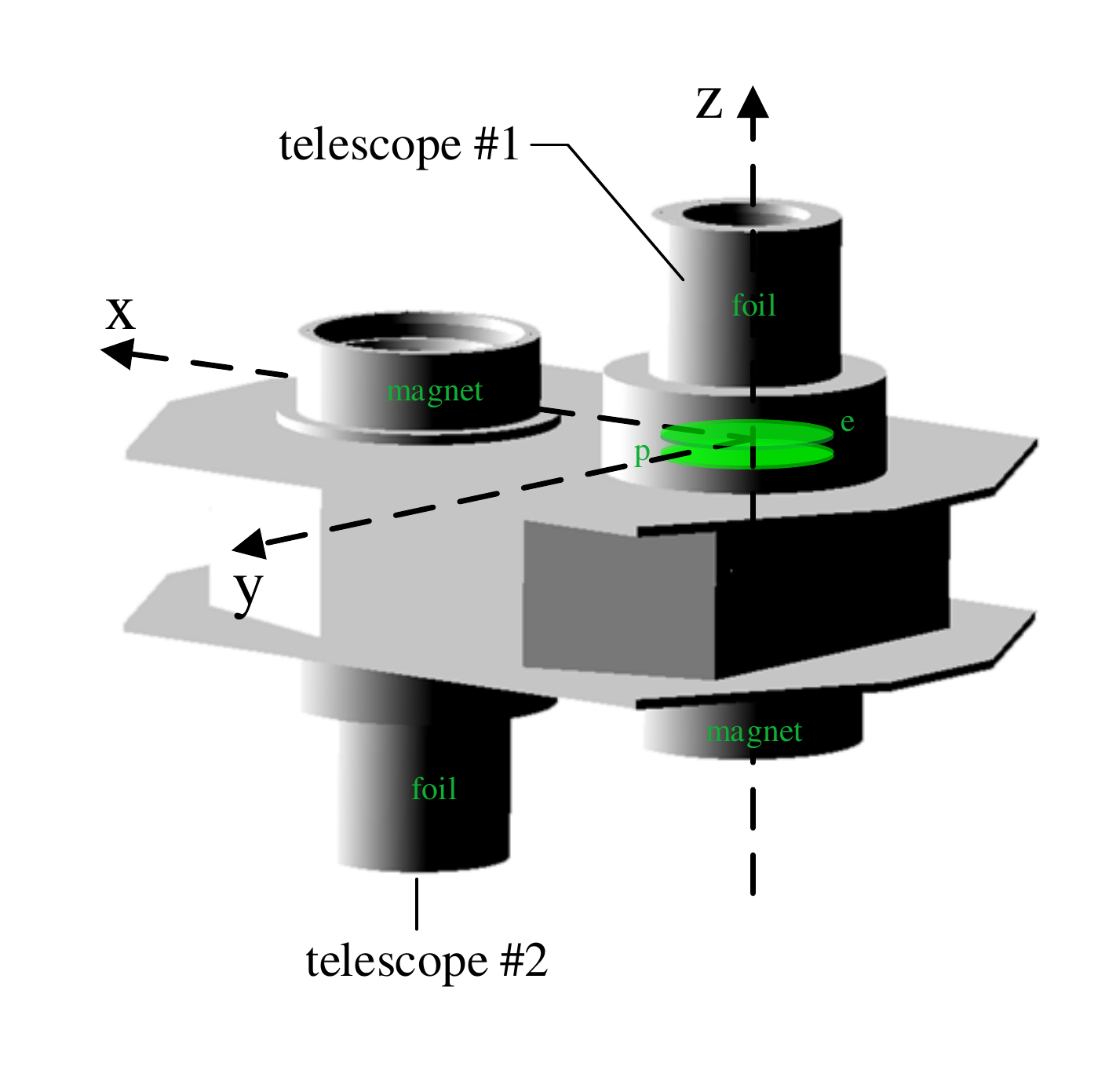}
\caption{SEPT detector model used in the GEANT4 simulation.}
\label{fig:sept-sketch-3d}
\end{figure}

\begin{figure}
\includegraphics[width=\columnwidth]{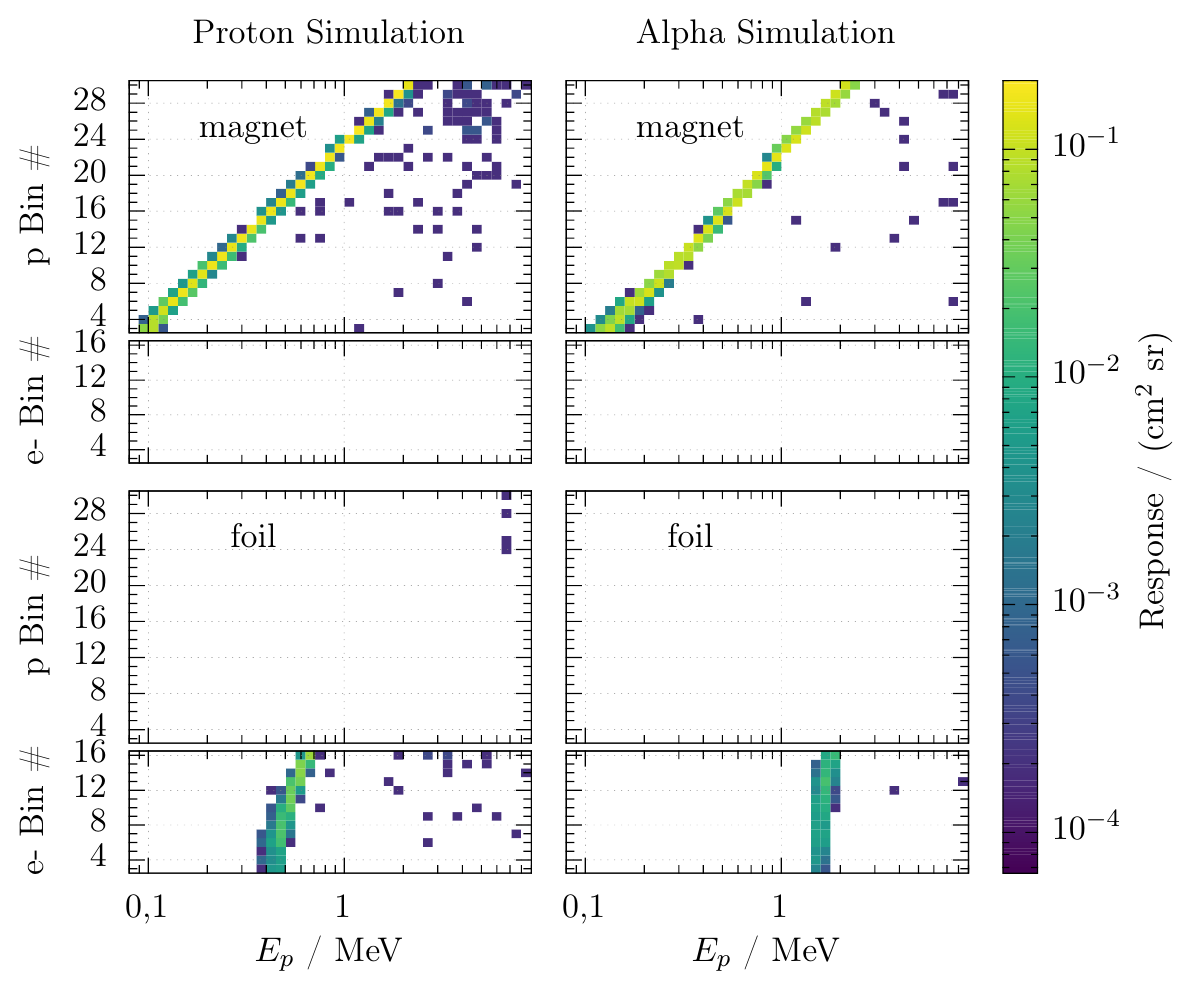}
\caption{From top to bottom: The left and right panels show the contribution of protons and alphas to the ion (p bins) and electron channels (e- bins) penetrating from the magnet and from the foil side, respectively.}
\label{fig:3}
\end{figure}

To determine this contribution we utilized the GEANT4 simulation of the SEPT as described in \citet{Mueller-Mellin-etal-2008}. Here we extended the simulation model by combining the two SEPT telescopes (providing measurements from both sides) and including the housing surrounding them. Figure~\ref{fig:sept-sketch-3d} displays the extended instrument model as used here. The green disks show the position of the two SSDs with the foil side pointing upwards and downwards for telescope 1 and 2 respectively. In order to determine the response of the instrument for protons and $\alpha$-particles respectively, particles with energies between 10~keV/nucleon and 10~MeV/nucleon that are injected isotropically by using a spherical source with a radius of 10~cm around SEPT have been tracked using the GEANT4 toolkit. Since the housing around SEPT shields the SSDs against protons and alpha particles with energies less than 10~MeV/nucleon we restricted the generation of particles within cones of 30$^\circ$ with respect to the instrument axis. As mentioned above the intensity excess due to the CIR event on 9~August 2011 is restricted to energies below 6~MeV/nucleon. Thus the simulations are sufficient to describe the energy spectra when subtracting the pre-event background. If we neglect the contribution of ions heavier than helium the flux in the channels can be described by a set of matrix equations:
\begin{equation}
\label{eq:counts}
\vec{C}
=
\bm{R}^p
\cdot \vec{I}^p
+
\bm{R}^\alpha
\cdot \vec{I}^\alpha
+
\bm{R}^e
\cdot \vec{I}^e
\end{equation}
with $\vec{C}$ the count rate vector consisting of $\vec{C}_p$ and $\vec{C}_e$ the count rate in the proton (p) and electron (e) channels. $\bm{R}^p$, $\bm{R}^\alpha$, and $\bm{R}^e$ are the responses of SEPT to incoming protons, $\alpha$-particles ($\alpha$) and electrons in the proton and electron channels with corresponding proton ($\vec{I}^p$), $\alpha$-particle ($\vec{I}^\alpha$) and electron ($\vec{I}^e$) fluxes. $\bm{R}^p$, $\bm{R}^\alpha$, and $\bm{R}^e$ have been computed in the energy range from $E_{low}^{j}$ and $E_{up}^{j}$ for $j=\lbrace 1,2, \cdots, m\rbrace $ dividing the energy range from 10~keV to 10~MeV in $m$ logarithmic equidistant intervals. Here $m=60$ is chosen to result in 20~bins per decade. Further we restrict our analysis to proton channels 3 to 30 (corresponding to energy deposits from 55 keV to 2.2~MeV) and electron channels 3 to 16 (55--425~keV) for which processed count rates are available and sufficiently unaffected by instrument noise. For simplicity we first assume that electrons are not accelerated by the CIR ($\vec{I}^e=0$). This boundary condition could be relaxed if the results are inconsistent with this assumption. The computed matrices for protons and $\alpha$-particles are displayed in Fig.~\ref{fig:3} left and right, respectively.

As expected a good correlation between the particle energy and the energy seen by the instruments is given for protons and $\alpha$-particles in the ion channels for particles penetrating through the magnet side (upper panels). Within the energy range of interest we get no signal in the electron channels. A similar situation is observed for particles penetrating through the foil. While the expected contribution to the electron channels can be found, there is no signal in the ion channels. We note the difference between protons and $\alpha$-particles: Following \citet{Mueller-Mellin-etal-2008} and assuming that the $\alpha$ and proton spectra are described by a power law in energy per nucleon with spectral index smaller than $-3$ and a constant ratio of about 10~\% the electron channels are dominated by protons and the ion channels are dominated by $\alpha$-particles \citep[for details see Fig.~14 in][]{Mueller-Mellin-etal-2008}. Fig.~\ref{fig:3} shows that the spectral shape and the $\alpha$ to proton ratio are well determined in the energy range from about 350 to 700~keV.

\begin{figure}
\includegraphics[width=\columnwidth]{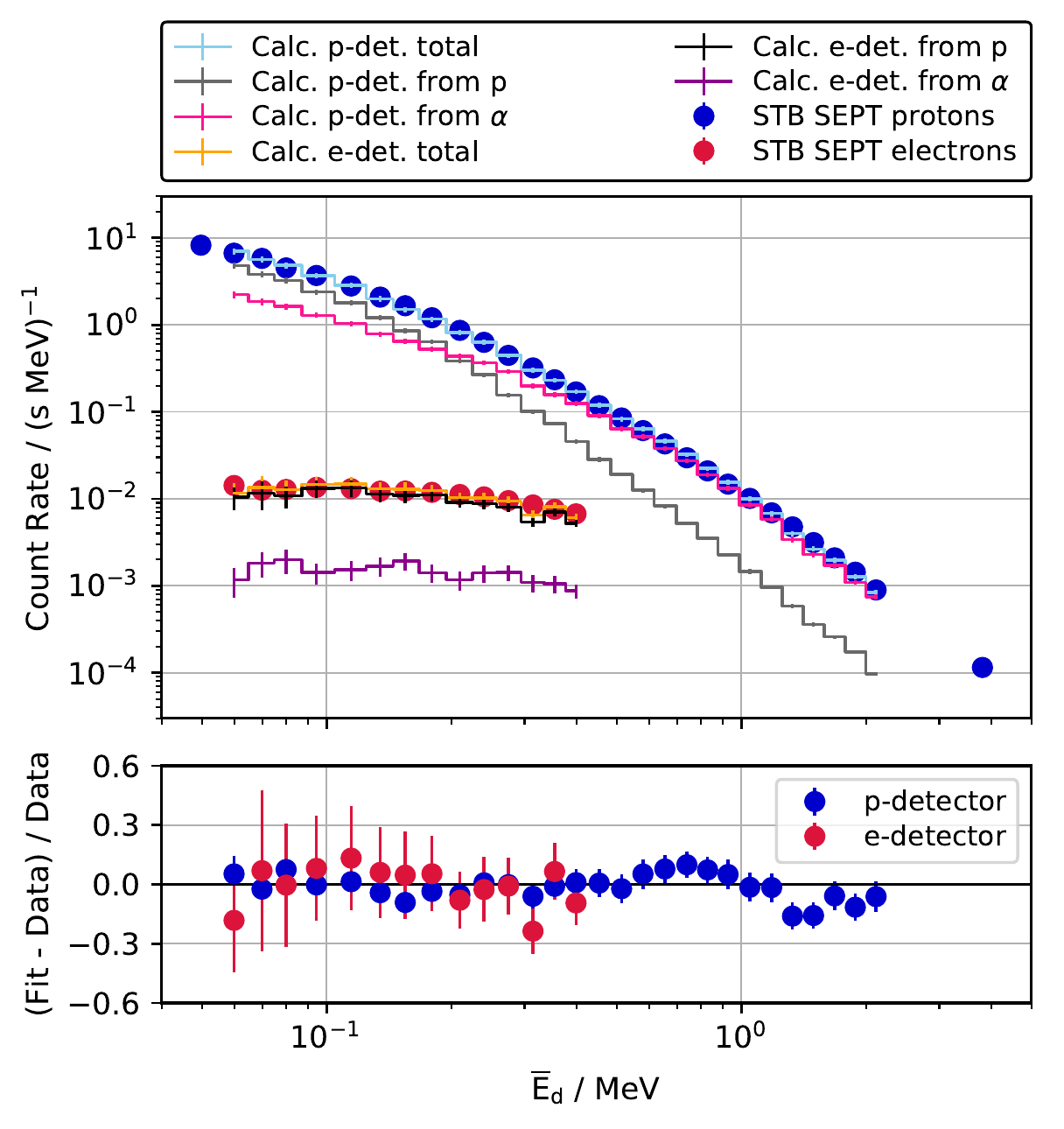}
\caption{Upper panel: Measured proton and electron count rates divided by the energy width of each channel are shown by the big dark-blue and red dots from 13 to 17~UT, respectively. The histogram with light-blue line gives the sum of the results of the fit for protons (light-magenta line) and $\alpha$-particles (gray line) in the ion channels. The lower orange line reflects the corresponding results for protons (black line) and $\alpha$-particles (dark-magenta line) in the electron channels. For completeness the ion flux from about 2.2 to 6.5~MeV is shown too. The lower panel displays the residuals for the ion and electron channels by the blue and red symbols, respectively.}
\label{fig:4}
\end{figure}

We will now apply our simulation results in order to determine the averaged proton and $\alpha$-particle spectra during 9~August 2011 from 13 to 17~UT. This time period includes the passage of the reverse shock of the CIR (see Fig.~\ref{fig:1}) and the measured particle flux is well above the pre CIR background. We note again that the measurements by LET indicate neither protons nor $\alpha$-particle acceleration above 10~MeV/nucleon. Therefore the computations were restricted up to 10 MeV/nucleon. 
From 13 to 17 UT we determine the measured SEPT proton and electron count rates. These count rates divided by the energy width of the corresponding channel are shown in Fig.~\ref{fig:4} by the blue and red dots for the proton and electron channels, respectively. Here, the count rates are plotted against the mean energy deposit $\overline{E}_\mathrm{d}$ of the channel. These count rates are used to ''solve'' the reduced equation (eq. (\ref{eq:counts}) with $\vec{I}^e=\vec{0}$)
\begin{equation} 
\label{eq:minimization}
\vec{C}
- 
\bm{R}^p
\cdot \vec{I}^p +
\bm{R}^{\alpha}
\cdot \vec{I}^\alpha =\vec{0}
\end{equation}
utilizing one Band function \citep{Band-etal-1993}
\begin{equation}
\label{eq:band-func}
\begin{aligned}
N(E) &= E^{\gamma_1} \exp(-\frac{E}{E_c})\\
     &\qquad \mathrm{for} \quad E < (\gamma_1 -\gamma_2)\cdot E_c \\
N(E) &= E^{\gamma_2} \left(\left[(\gamma_1-\gamma_2)\cdot E_c\right]^{\left(\gamma_1-\gamma_2\right)}\cdot \exp(\gamma_1-\gamma_2)\right)\\
     &\qquad \mathrm{for} \quad E \geq \left(\gamma_1-\gamma_2\right)\cdot E_c
\end{aligned}
\end{equation}
to constrain the proton and $\alpha$-particle spectra $\vec{I}^p$ and $\vec{I}^\alpha$ as following:
\begin{align}
I^{p}(E) &= N(E) \cdot I_0\,,\label{eq:Ip}\\
I^\alpha(E) &= I^{p}(E) \cdot \alpha\,.\label{eq:Ia}
\end{align}
These constrained spectra leave the $\alpha$-particle to proton ratio $\alpha$, $I_0$, $\gamma_1$, $\gamma_2$, and $E_c$ as free parameters in eq. (\ref{eq:minimization}). $E_c$ and $E$ are given in energy per nucleon. $\gamma_1$ and $\gamma_2$ give the slope of the spectrum at low and high energies, respectively \citep[see also ][and references therein]{Mewaldt-etal-2012}. The vector elements of $\vec{I}^p$ and $\vec{I}^\alpha$ are mean values in the energy interval of the corresponding energy bin calculated from eq. (\ref{eq:Ip}) and (\ref{eq:Ia}) using the Band function. Since solving eq.~(\ref{eq:minimization}) under this assumption is impossible one normally determines the absolute value and varies the free parameters in order to minimize it.
Here we perform the minimization with a variant of the minimum chi-square method as used in \cite{Koehler-etal-2011} utilizing the following equation:
\begin{equation}
\label{eq:minimizer}
\mathrm{min} \sum_i \left(\frac{\sum_j \left( r_{ij}^p I_j^p + r_{ij}^\alpha I_j^\alpha \right) - c_i}{c_i}\right)
\end{equation}
where $r_{ij}^p$, $r_{ij}^\alpha$, $I_j^p$, $I_j^\alpha$ and $c_i$ are the components of the matrices $\bm{R}^{p}$, $\bm{R}^{\alpha}$ and the vectors $\vec{I}^p$, $\vec{I}^\alpha$, $\vec{C}$ from eq. (\ref{eq:minimization}), respectively. We perform this minimization using the implementation of the Levenberg-Marquardt algorithm in the \texttt{lmfit} module \citep{python-lmfit} for Python 2.7. The output of the minimization procedure are the fitted proton and $\alpha$ spectra $\vec{I^p}$ and $\vec{I^\alpha}$ as well as the corresponding SEPT count rates calculated from eq.~(\ref{eq:counts}). The contribution of each component is also given by the same equation.

The resulting count rates divided by the energy width of each SEPT channel are displayed in Fig.~\ref{fig:4} by the light-blue and orange lines. The contribution by protons and $\alpha$-particles to the ion and electron channels are given by the light-magenta and gray as well as by the black and dark-magenta lines, respectively. The corresponding parameters resulting from the minimization are an $\alpha$-particle to proton ratio of 16.9~\%, $I_0=1.24\cdot 10^{4}$~/~(cm$^2$~s~sr~MeV/nuc.), $E_c = 79$~keV/nuc. and spectral indices $\gamma_1=-0.94$ and $\gamma_2=-3.80$. The lower panel shows the residuals for the ion (blue symbols) and electron channels (red symbols). The error bars were calculated using the statistical error of the simulation and from the data. As expected the high energy ion channels are dominated by $\alpha$-particles and the electron channels are entirely dominated by protons. Here we conclude that the measurements are well represented by the proton and $\alpha$-particle contribution, thus leaving no indication for an additional electron component. We note that the correlation between $\sim 100$~keV electrons and $\sim 100$~keV ions at the onset of this ion channel at around 8:00 UT is negligible (see Fig. \ref{fig:1}), so that the existence of pinholes in the parylene foil of the SEPT, which would have an effect on the instrument response, is unlikely.

\begin{figure}
\includegraphics[width=\columnwidth]{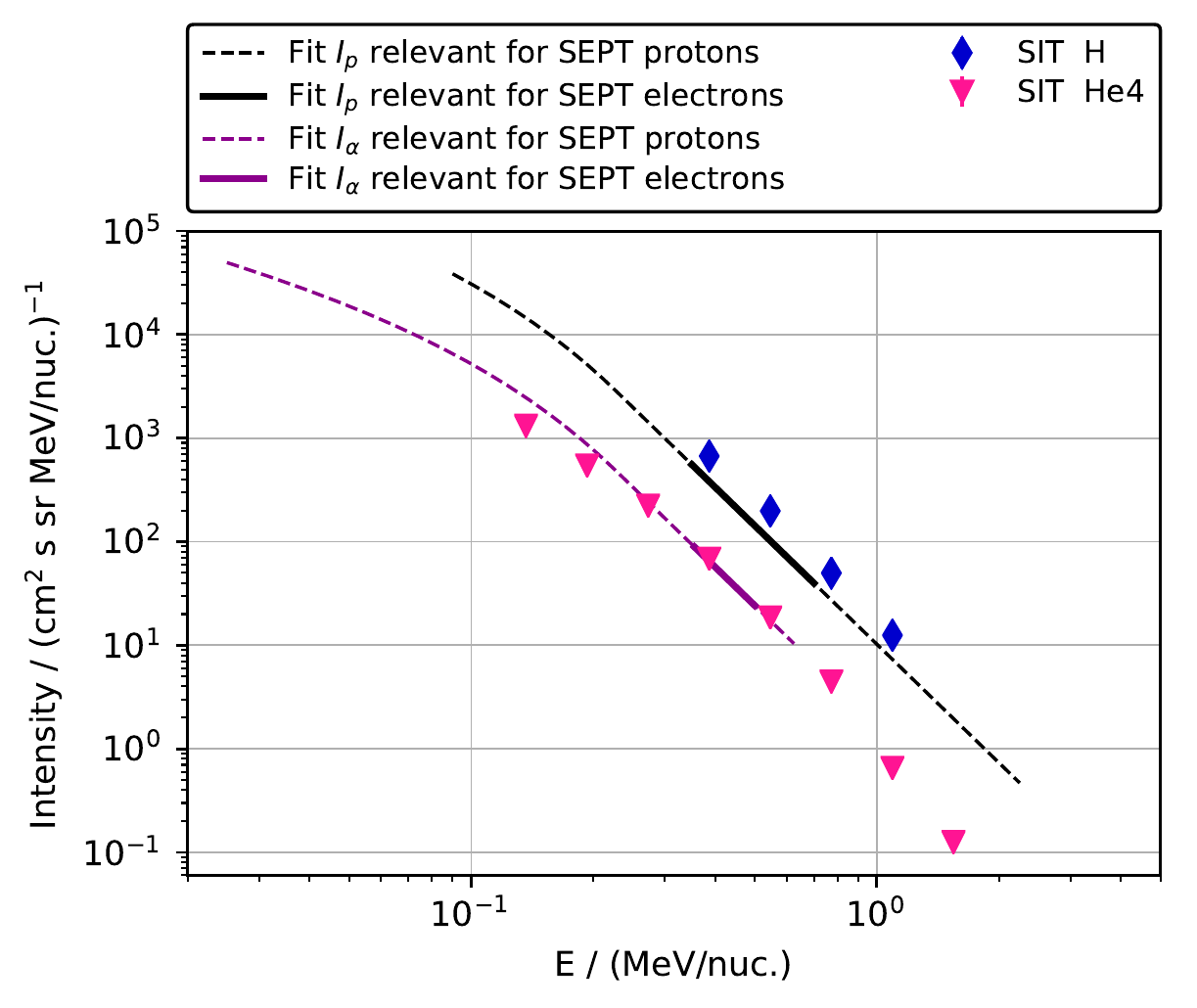}
\caption{Proton (blue symbols) and helium (magenta symbols) energy spectra during the time period from 13 to 17 UT as measured by the STEREO SIT \citep{Mason-etal-2008}. In comparison the black and dark-magenta lines show our results from analyzing SEPT measurements. Here, the solid lines represent the part of the fitted spectra which contributes to all SEPT channels. The part marked by the dashed lines contributes to the proton channels only.}
\label{fig:5}
\end{figure}

The proton and $\alpha$-particle spectra that result from our analysis are displayed by the black and dark-magenta lines in Fig.~\ref{fig:5}. While the solid segments of the lines in Fig.~\ref{fig:5} represent the energy range of each spectra which contributes to both the SEPT proton and electron channels, the energy range marked by the dashed segments contributes to the proton channels only (see the responses in Fig~\ref{fig:3}). Thus, the most confident fit can be found in the solid lines. For comparison measured proton and helium spectra by the the Suprathermal Ion Telescope \citep[SIT,][]{Mason-etal-2008} aboard STB during the same time period are shown by blue and magenta symbols. The determined $\alpha$-spectra are in good agreement. Moreover, the spectral shape of the determined proton spectrum, given by $\gamma_2 = -3.80$, matches the data very well. Only the $\alpha$-particle to proton ratio of 16.9~\% resulting from the minimization is overestimated compared to a ratio of 10~\% determined from the SIT data.

\section{Summary}
We analyzed the CIR energetic particle event that was observed by SEPT aboard STB on 9~August 2011 that was interpreted by \citet{Xu-etala-2017} as a signature of an SEP event that caused electron and proton flux increases near Earth at L1. Here we showed that the apparent electron flux enhancement at STB can be fully understood by the ion (mainly proton) contamination in the electron channels due to the high ion flux. The reason for such contamination lies in the measurement principle of the magnet foil technique used by SEPT. We utilized a GEANT4 simulation of SEPT in order to compute the response function to protons and $\alpha$-particles of the instrument. Under the assumption that no electrons contribute to the flux from 13 to 17~UT during the passage of the reverse shock of the CIR and that the proton and helium spectra are given by a Band function in energy per nucleon with a constant $\frac{\alpha}{p}$-ratio we could infer the proton and helium spectra that were measured by SEPT during that period. The obtained fluxes are in good agreement with the one measured by SIT aboard STB, leaving no room for an additional electron component. Our study supports the one by \citet{Dresing-etal-2016} showing that the near-relativistic electron acceleration by CIR shocks at 1~au is not very common. A critical data analysis is recommended and has to be applied before interpreting ''electron measurements'' obtained by utilizing the magnet foil technique like SEPT.

\begin{acknowledgements}
This paper uses data from the Heliospheric Shock Database, generated and maintained at the University of Helsinki. The STEREO/SEPT project is supported under grant 50OC1702 by the Federal Ministry of Economics and Technology on the basis of a decision by the German Bundestag. Reproduced with permission from Astronomy \& Astrophysics, \copyright ESO
\end{acknowledgements}

\bibliographystyle{aa}
\bibliography{references}


\end{document}